# ASTROPHYSICS: Most distant cosmic blast seen

Bing Zhang

*The most distant γ-ray burst yet sighted is the earliest astronomical object ever observed in cosmic history. This ancient beacon offers a glimpse of the little-known cosmic dark ages.*

In this issue, two papers[1,2] report the discovery a γ-ray burst (GRB) at a redshift of about 8.2. It is the highest redshift — or equivalently the most distant — astronomical object ever detected in the Universe. For comparison, the highest redshift recorded so far for galaxies is about 6.96 (ref. 3). For quasars — extremely bright galactic nuclei powered by supermassive black holes — the record holder is an object at 6.48 (ref. 4).

Tanvir *et al.*[1] (page 1254) and Salvaterra *et al.*[2] (page 1258) measured a concordant redshift for GRB 090423 on the basis of observations of its fading afterglow emission at different times after the initial burst of γ-rays. In astronomy, a larger distance (or redshift) corresponds to an earlier time in cosmic history, because it takes longer for a farther-away photon, which travels at finite speed, to reach Earth. Moreover, because the Universe is expanding, the wavelength of electromagnetic radiation emitted by a distant object is stretched to be longer and redder (redshifted) on its course to Earth. The farther away the source, the more the Universe has expanded, and therefore the higher the source's redshift. The redshift measured for GRB 090423 means that the burst occurred at a time when the Universe was about nine times smaller than it is today — putting the timing of the event at about 630 million years after the Big Bang.

GRBs are the most violent explosions in the Universe. They are believed to be associated with the formation of stellar-sized black holes or rapidly rotating, highly magnetized neutron stars during cataclysmic events such as the collapse of a massive star or the coalescence of two compact stellar objects. The reason why GRBs can outshine galaxies and quasars, which are much more massive, to become the redshift record holders (Fig. 1) is threefold.

First, thanks to their extremely high speed (99.9995% of the speed of light), their maximum luminosity is extremely high, dwarfing that of galaxies and quasars by many orders of magnitudes.

Second, whereas galaxies and quasars look progressively dimmer at higher redshifts (both because of a larger distance from Earth and an intrinsic smaller mass at an earlier cosmic time), the apparent brightness of GRBs and their infrared afterglows (which hold the key to redshift identification) do not decrease significantly with increasing redshift. This is due to a concerted combination of two effects, the *k*-correction and the time-dilation effect[5].

Finally, whereas bright galaxies and quasars become rarer as redshifts rise above 7, theoretical models suggest that massive stars, thought to be the progenitors of high-redshift GRBs, can form much earlier[6]. As a result, GRBs can be more easily detected at higher redshifts, and may hold the key to illuminating the cosmic 'dark ages' (Fig. 2).

Although satellites such as Swift[7] have been successful in detecting high-redshift GRBs, characterizing these objects and measuring their redshifts is a different matter. To do so, requires catching their rapidly fading afterglows and identifying their spectral signatures. The main signature is a fall in emissions, caused by intervening hydrogen gas clouds along the light path, bluewards of the Lyman-α resonance line of hydrogen. For high-redshift GRBs, this feature is best observed with near-infrared observations, because the original ultraviolet light of such GRBs is redshifted into the observable near-infrared window.

For GRB 090423, Tanvir et al.[1] observed the near-infrared afterglow, starting about 17.5 hours after the burst, using the European Southern Observatory 8.2-m Very Large Telescope in Chile. Meanwhile, Salvaterra et al.[2] observed the burst starting from about 14 hours after it occurred using the 3.6-m Italian Telescopio Nazionale Galileo in Spain. Both teams discovered a clear break in emissions at wavelengths shorter than about 1.1 micrometres, which corresponds to the Lyman-α line frequency at the derived redshift.

The authors' discovery[1,2] opens up the exciting possibility of studying the cosmic 're-ionization' epoch, and the preceding dark ages, using GRBs. In its spectrum, particularly in the absorption 'damping wing', a GRB carries information about the fraction of neutral gas in the intergalactic medium (IGM)[4], and so the IGM reionization state, at the cosmic time at which it occurred. Studying a number of GRBs spread over a range of redshifts would allow one to map out the course of the re-ionization process throughout cosmic time.

Realistically, two factors seem likely to hinder such prospect. First, afterglow observations suggest that there is a large amount of neutral gas in the interstellar medium (ISM) surrounding GRBs. In some cases the contribution of the ISM to GRB-light absorption dominates over that of IGM absorption, making it difficult to extract information about cosmic re-ionization from the data[8]. But cosmological simulations suggest that, at high redshifts, this ISM effect should decrease with increasing redshift[9], a trend that seems to be supported by observations of the second-highest-redshift burst, GRB 080913 (ref. 10). More high-redshift GRB data are needed to verify or disprove this prediction.

Second, afterglows fade rapidly with time. High-resolution spectroscopy is therefore needed at early (post-burst) epochs to catch them when they are still bright. In the case of GRB 090423, high-resolution spectra[1,2] were taken from about 14 hours after the burst, when the afterglow had already faded considerably. The spectra presented by Tanvir *et al.*[1] and Salvaterra *et al.*[2] can therefore only serve the purpose of redshift identification. To make further progress in studying the high-redshift Universe with GRBs, prompt alerts on high-redshift GRB candidates would be desirable. It is hoped that this could be achieved in the future by GRB space missions such as SVOM, JANUS, and EXIST, or by ground-based, large near-infrared robotic telescopes.

That said, GRB 090423 still provides much information about the high-redshift Universe, and the mechanisms that underlie GRB formation at that cosmic epoch. Salvaterra and colleagues[2] notice that the burst detection may suggest that star-formation rate at high redshift is anomalously high, or that the GRB luminosity function (the relative number of GRBs that have a certain luminosity) evolves with redshift. Tanvir and colleagues[1], however, point out that its detection is consistent with the high-redshift star formation rate predicted by some theoretical models[11]. More data are needed to solve the authors' discrepant interpretations of the detection of GRB 090423.

Moreover, both studies[1,2] report that the X-ray and near-infrared afterglows of the burst are not very different from those of nearer GRBs, suggesting that progenitor stars similar to those of their more recent counterparts already existed at cosmic times as early as 630 million years after the Big Bang. Detection of radio afterglow and afterglow modeling[12] suggest that the circumburst medium of GRB 090423 is also similar to that of its nearby cousins. Taken together, all of these observations[1,2,12] indicate that the progenitor of GRB 090423 is not one of the first-generation stars, which, unlike their present-day analogues, are believed to be much more massive and metal poor (containing only hydrogen and helium)[6].

Finally, the (rest-frame) duration of GRB 090423 is slightly longer than 1 second. Given its redshift, this is an unexpected property, but one that is shared by GRB 080913, and that implies that the burst may fall into the short-lived category. However, several arguments[13] suggest that both GRB 090423 and GRB 080913 are related to the collapse of massive stars, rather than the merging of compact objects that is believed to power nearby, short-lived GRBs[14].

The apparently short durations of these two bursts may be due to an observational selection effect. A more intriguing possibility, which future observations could test, is that this is related to the properties of GRB progenitors at increasingly higher redshifts, which seem to produce intrinsically shorter bursts. In any case, thus far, the indication is that GRB duration alone is no longer necessarily the crucial criterion to discern the physical nature of GRBs[13].

Bing Zhang is in the Department of Physics and Astronomy, University of Nevada, 4505 Maryland Parkway, Las Vegas, Nevada 89154-4002, USA.

e-mail: zhang@physics.unlv.edu

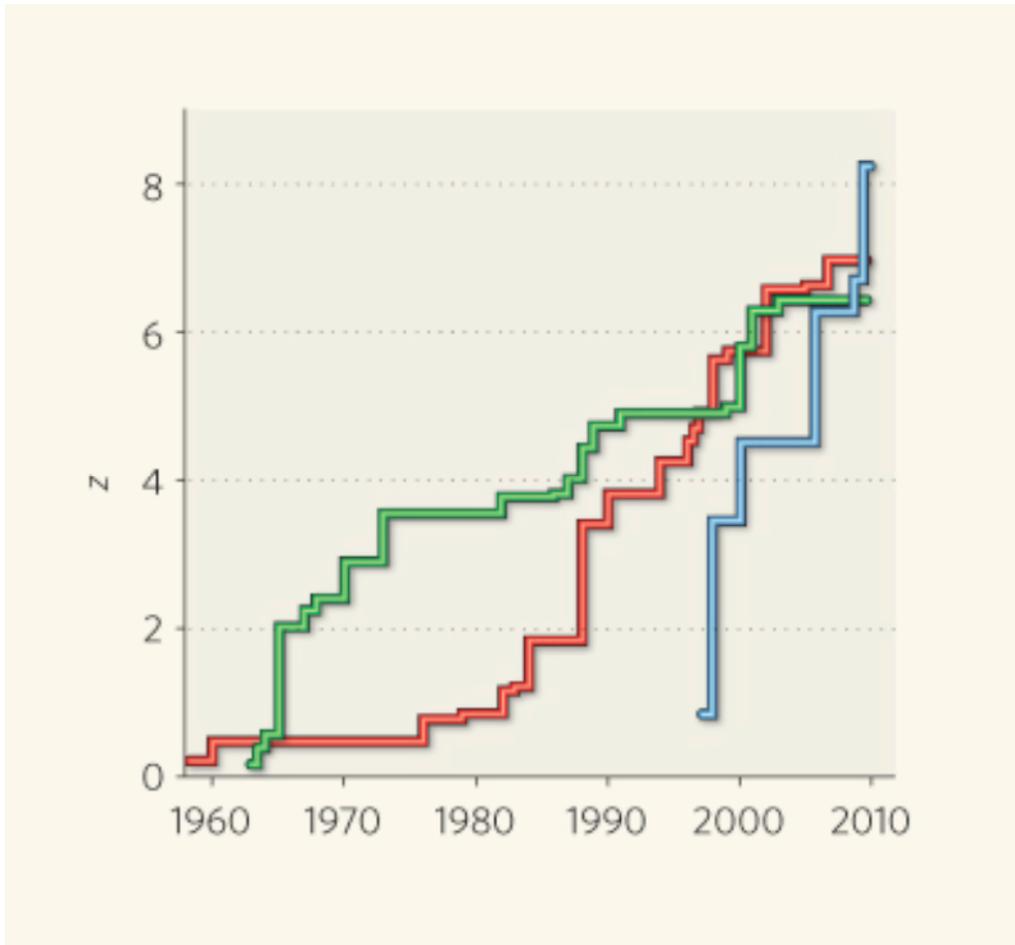

**Figure 1 Redshift ladder.** Timeline of redshift record-breaking for three classes of astronomical object: galaxies (red), quasars (green) and γ-ray bursts (blue). The measurement of γ-ray-burst redshift began much later than that of galaxies and quasars, but since the first discovery it has seen a much faster pace of redshift record-breaking. (Courtesy of Nial Tanvir.)

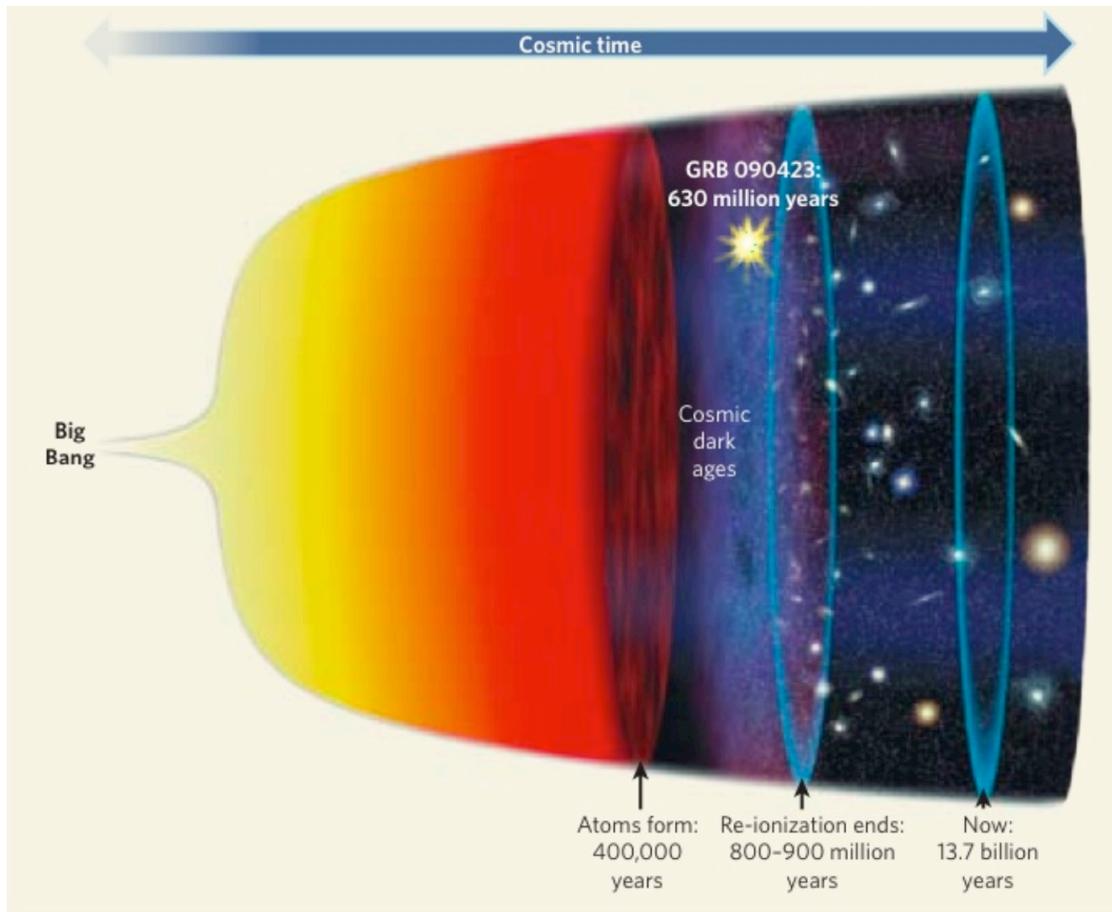

**Figure 2 The cosmic dark ages and GRB 090423.** After the Big Bang, the Universe cools rapidly while expanding. About 400 thousand years after this event, free electrons and protons combine to form neutral atoms, leaving a bath of background radiation that currently shines in the microwave part of the electromagnetic spectrum. Thereafter, the Universe remains neutral, until the first stars and galaxies light up at a later epoch. Photons emitted by these objects knock electrons out of atoms and 're-ionize' the Universe. Studies of the most distant galaxies and quasars suggest that the re- ionization process was completed around 800–900 million years after the Big Bang, but no information is available about the cosmic 'dark ages'. Observations of γ-ray bursts such as GRB 090423 (refs 1 and 2), which occurred about 630 million years after the Big Bang, offer a glimpse of the cosmic dark ages. (Adapted from ref. 15)